# On the Unambiguous Distance of Multi-Carrier Phase Ranging with Random Hopped Frequencies

Peng Liu, *Student Member, IEEE,* Wangdong Qi, *Member, IEEE,* Yue Zhang, and Li Wei

***Abstract*—Range estimation ambiguity is a fundamental problem in multi-carrier phase ranging (MPR) method since it limits the measurable distance of MPR. In favor of ambiguity resolution, it is desirable to extend the unambiguous range (UR). It is well known that the UR is disproportional to the frequency step size $\Delta f$ under the commonly used *linearly spaced frequency (LSF)* configuration. Although a smaller $\Delta f$ leads to a larger UR, the number of frequencies $N$ should also be increased since the occupied bandwidth $N\Delta f$ determines the ranging resolution. In this paper, we make a first attempt to study the UR under *randomly spaced frequency (RSF)* configuration which employs only a small number of, say $M$, random frequencies out of a large LSF set. It is proved that the upper bound of UR $\Lambda_{up}$ under RSF configuration is equivalent with that under original LSF one as long as those $M$ frequencies are relatively prime. Using growth estimation technique from analytic number theory, we are inspired to find that this condition could be easily met with high probability when $M > 10$, several orders less than $N$. This theoretical conclusion is highly supported by numerical simulations. However, it may be over-optimistic in practical scenario since there would exist multiple pseudo URs within $\Lambda_{up}$ due to phase noise. Fortunately, the theoretical UR under RSF configuration is much more resilient against phase noise than LSF configuration, because the randomness of frequencies disrupts the paces of pseudo URs and makes them more distinguishable from the true UR. The significance of our work lies in that it facilitates the application of MPR in large scale radio positioning/navigation system since the measurable distance of MPR could be highly scaled by only employing a small number of random frequencies rather than a large frequency set.**

## I. Introduction

MULTI-CARRIER phase ranging (MPR) has been commonly applied for radio navigation [1,2,3], telemetry [4], RADAR [5], smart sensor localization [6] and wireless network security [14]. In MPR, range is estimated from phase shift between the received and reference signals. When the distance to be measured is larger than carrier wavelength, MPR has to solve the problem of range estimation ambiguity. Thus, the measurable distance of MPR should always be smaller than its unambiguous range (UR) $\Lambda$, which is also referred as lane width in OMEGA [1] and carrier phase GPS (CP-GPS) [3], within which unique range estimation exists. The determining factor for $\Lambda$ is the carrier frequency set $\{f_i\}$. So far, the relationship between $\Lambda$ and $\{f_i\}$ under the linearly spaced frequency (LSF) set configuration, as shown in Fig. 1(a), has been well studied. A basic conclusion is that $\Lambda$ is disproportional with the frequency step size $\Delta f$ [2, 5, 6]. For example, given that UR is no less than 30 meters, then $\Delta f$ must be confined to $10MHz$; and if UR extends to be 300 meters, then $\Delta f \leq 1MHz$. In favor of ambiguity resolution, the UR should be as large as possible. Intuitively, a smaller $\Delta f$ leads to a larger $\Lambda$. Nevertheless, the number of measurement frequencies $N$ should also be increased since the occupied bandwidth $N\Delta f$ determines the ranging resolution [5]. Therefore, a trade-off arises between the UR and the real-time performance of MPR. For example, a regional radio navigation system in [2] tries to extend the UR to miles level by employing a large frequency set with $N$=192. The time-consuming measurement process takes as long as 6 seconds, making MPR unsuitable for positioning/navigating dynamic targets. Obviously, the UR could not be arbitrary large considering the constraint of real time requirement.

In this paper, we provide a new method to extend the UR of MPR without sacrificing its real time performance by employing a randomly spaced frequency (RSF) set. To be specific, the frequency step size $\Delta$f is firstly determined by the requirement on UR. Given the occupied bandwidth B, we have a large candidate frequency set $\mathbb{N}$. Then a small fraction of, say $M$, frequencies are randomly chosen from $\mathbb{N}$ to form the measurement set, as shown in Fig. 1(b). We make a first attempt to analyze the relationship between UR and $\{f_i\}$ under the condition that the measurement frequencies are random variable. This configuration is of particular interest when MPR method is implemented with Frequency Hopping Spread Spectrum (FHSS) signal. FHSS signal quickly switches carrier frequency following a pseudo random sequence and attains the fame of strong anti-jamming capability. Although MPR with random hopped frequencies has been adopted in continuous wave RADAR, the range estimation is obtained by rearranging those frequencies in a rising manner after phase shifts at all N frequencies are measured. That is, measurement process follows a random hopped frequency pattern while ambiguity resolution is just like that under the LSF configuration [5].

However, $\Lambda$ under RSF configuration becomes a random variable and great difficulty arises when we try to describe its stochastic property. Alternatively, we turn to focus on the upper bound of $\Lambda$ under the RSF configuration. Firstly, it is proved that the upper bound of UR under RSF configuration $\Lambda_{up}$ is the same as that when employing the whole candidate frequency set $\mathbb{N}$ as long as those $M$ frequencies are relatively prime. Using growth estimation technique from analytic number theory, we obtain an elegant expression to describe the probability $P_M$ that the UR achieves $\Lambda_{up}$. It is inspiring to find that this probability would approach to 1 very closely when $M$ is slightly larger than 10. This conclusion indicates that the UR could have been extended to be sufficient large by employing any small fragment of a random hopped frequency sequence, and therefore, large measurable distance and good real time performance could be achieved at the same time. Numerical simulation highly corresponds with this theoretical conclusion.

A very interesting phenomenon lies in that if those $M$ frequencies are chosen linearly from candidate set $\mathbb{N}$, it is not surprising to get a relatively prime measurement set. We reveal that there exist multiple pseudo UR in this scenario which is only a fractional of $\Lambda_{up}$. Considering the impact of phase noise, those pseudo UR is no longer distinguishable from the theoretical one. On the contrary, the UR under RSF configuration is shown to be more resilient against phase noise through extensive simulation. So the theoretical upper bound is also achievable in practical radio navigation/positioning system.

The rest of this paper is organized as follows. The system model is described in section II and a thorough proof on the upper bound of UR as well as the probability that it could be obtained is given in section III. In section IV, numerical simulations are conducted to verify the theoretical conclusion without phase noise, and then Monte-Carlo simulation results are provided to show the robustness of UR against phase noise under the LSF and RSF configurations. Finally, we reach the conclusion in section V.

## II. System model

### A. Ambiguity Resolution in Multi-carrier Phase Ranging

Range estimation in MPR is obtained through measuring phase shift between reference signals and received signals at multiple frequencies. Denote the reference signal as $x(t) = A_t \cos(2\pi f_i t)$ and the received signal is the addition of delayed signal and channel noise $y(t) = B_t \cos(2\pi f_i(t - d/C)) + n(t)$ where $A_t$ and $B_t$ are signal amplitudes, $d$ is the distance between the transmitter and the receiver, $C$ is signal transmission speed, $n(t)$ is the channel noise and $\mathrm{i} = 1, \cdots \mathrm{N}$ where N is the number of carriers. Here we only consider the one way ranging method. The phase shifts between reference signals and received signals are proportional to the distance $d$ given the constant of signal propagation speed, that is,

$$\varphi_i = 2\pi f_i \frac{d}{C} \quad (mod\ \ 2\pi)\ . \tag{1}$$

Due to the modulo operation, there exist ambiguity when deducing d from phase measurements, that is,

$$\mathrm{d} = n_i\lambda_i + \frac{\varphi_i}{2\pi}\lambda_i\ , \tag{2}$$

where $\lambda_i = C/f_i$ is the carrier wavelength and $n_i$ is the whole cycle ambiguity. Since $\varphi_i$ is contaminated by noise, those equations in (2) could not be solved analytically. Ambiguity may be solved by phase unwrapping [*] or other estimators. Here we utilize the least square estimator developed in [6] which is represented as follows:

$$\hat{\mathrm{d}} = \arg\ \min_d F(d)\ , \tag{3}$$

where $\mathrm{F(d)} = \sqrt{\sum(\varphi_i - 2\pi\frac{d}{\lambda_i})^2}$ is the discrepancy function. The value of $\mathrm{F(d)}$ is also called discrepancy error of estimation $d$.

**Definition 1**. The Unambiguous Rage of MPR is defined as $\Delta\mathrm{d}$ which meets the following criterion:

$$\mathrm{prob}(\hat{\mathrm{d}} = \mathrm{d}_0) = \mathrm{prob}(\hat{\mathrm{d}} = \mathrm{d}_0 + \Delta\mathrm{d})$$

where $\mathrm{prob}(\cdot)$ is the probability that that range estimate takes a specific value.

### B. Carrier Frequency Set Configuration

The unambiguous range (UR) $\Lambda$ of MPR is determined by the carrier frequency set $\{f_i\}$, denoted as $\mathbb{N}$. In the linearly spaced frequency (LSF) set configuration, as shown in Fig. 1(a), $\{f_i\}$ is an arithmetic sequence. Under the LSF configuration, it is well known that UR is disproportional with frequency step, that is, $\Lambda = \mathrm{C}/\Delta\mathrm{f}$. So, firstly given a certain $\Lambda$, we could get the corresponding $\Delta\mathrm{f}$. Then given the occupied bandwidth $\boldsymbol{B}$, which is the dominating factor of the ranging resolution, the number of carriers $\mathrm{N} = \frac{\mathrm{B}}{\Delta\mathrm{f}} + 1$ is also determined.

Another configuration which is of great interest in this paper is the randomly spaced frequency (RSF) set. Suppose each frequency $f_i \in \mathbb{N}$ could be represented as $f_i\ =\ k_i f_0$ where $f_0$ is the basic resolution of radio frequency synthesizer, $k_i$ is the value of frequency register. It should be noted that the frequency step $\Delta f$ in LSF may be a multiple of $f_0$. Let $\mathbb{K}$ denote the set of positive integers $\{k_i\}$, we have $K_{min} \le k \le K_{max}$ where $K_{min}$ and $K_{max}$ correspond to the lower and upper boundary of available bandwidth respectively. Then $M$ integers are randomly chosen from $\mathbb{K}$ to form a much smaller measurement set, as shown the red lines in Fig. 1(b). Particularly, the frequency set $\mathbb{N}$ under RSF configuration might be either continuous or discontinuous with the following considerations: (a) in the dynamic spectrum access paradigm enabled by cognitive radios, the distribution of free spectrum is often discontinuous [8]; and (b) FHSS signal could readily utilize discontinuous bandwidth to further improve its accuracy [9]. So we have $\mathbb{K} \subseteq [K_{min},\ K_{max}]$ since some partial bandwidth may have already been occupied. Further suppose $\mathbb{K}$ is separated into $L$ segments with $N_l$ be the number of available frequencies in the $l$th segment. Then the norm of $\mathbb{K}$ is $\mathrm{N} = \sum_{l=1}^{L} N_l \le N_0$ where $N_0 = \mathrm{B}/f_0 + 1$ is the maximum number of available frequencies within bandwidth $\boldsymbol{B}$.

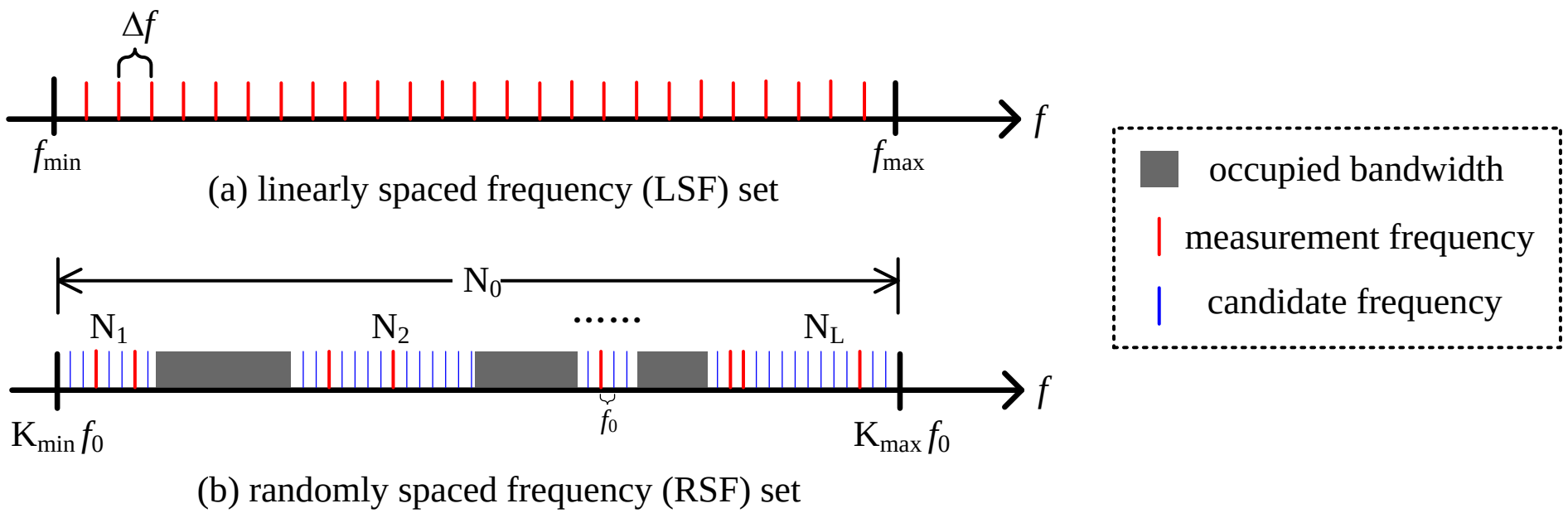

Figure 1. Measurement frequency set of MPR under LSF and RSF configurations

Table 1. Terms used in the description of LSF and RSF configuration

| Term | Definition | Note |
|---|---|---|
| $\Lambda$ | The unambiguous range of MPR | |
| $\boldsymbol{N}$ | The number of carriers in the LSF set | A large integer |
| $\boldsymbol{M}$ | The number of carriers in the RSF set | A small integer |
| $\boldsymbol{B}$ | Occupied bandwidth | |
| $\boldsymbol{L}$ | The number of sub-bands within $\boldsymbol{B}$ | A small integer |
| $\boldsymbol{N_l}$ | The number of available carriers in the $l$-th sub-band | |
| $f_0$ | Resolution of radio frequency synthesizer | May be as low as 1Hz |
| $\Delta f$ | Frequency separation in the LSF set | Integer times of $f_0$ |
| $f_i$ | The i-th carrier frequency | $i=1,\ldots,N$ |
| $k_i$ | The register value corresponding to $f_i$ | $k_i = f_i / f_0$ |
| $\mathbb{N}$ | Measurement frequency set of $\{f_i\}$ | |
| $\mathbb{K}$ | Frequency register set of $\{k_i\}$ | |

## III. Unambiguous Range under RSF Configuration

The UR under RSF configuration is a random variable related with those measurement frequencies. Because of the nonlinearity of range estimation, great difficulties arise to describe the stochastic property of UR. Alternatively, in this section we would firstly give the upper bound of UR $\Lambda_{\mathrm{up}}$ and then analyze the probability that $\Lambda_{\mathrm{up}}$ could be obtained under RSF configuration.

### *A. The Upper Bound of UR*

Theorem 1 describes the general relationship between UR and the corresponding measurement frequency set, independent of specific configuration.

**Theorem 1**. The UR $\Lambda$ of MPR could be expressed in terms of measurement frequency set $\{f_i\}$ as follows:

$$\Lambda = \frac{C}{\kappa f_0}, \tag{4}$$

where $\kappa$ is the greatest common divisor (GCD) of $\{k_i\}$.

***Proof:*** Theoretically, it has been proved that $\Lambda$ is the least common multiple (LCM) of all carrier wavelengths $\{\lambda_i\}$ [10]. Since $\lambda_i = C/k_i f_0$, we turn to find the LCM of $\{1/k_i\}$. Denote $1/\kappa'$ as one of the common multiples of $\{1/k_i\}$, then $1/\kappa'$ is divisible by $1/k_i$ if and only if $k_i$ is divisible by $\kappa'$, which indicates that $\kappa'$ is the common divisor of $\{k_i\}$. If $\kappa'$ is the GCD of $\{k_i\}$, i.e. $\kappa$, then $1/\kappa$ is the LCM of $\{1/k_i\}$. Therefore, $\Lambda$ could be expressed as $\frac{C}{\kappa f_0}$.

Based on theorem 1, we could easily prove the following lemma.

**Lemma 1**. The upper bound of UR is $\Lambda_{up} = \frac{C}{f_0}$ and it could be obtained if and only if those $\{k_i\}$ out of $\mathbb{K}$ are relatively prime.

***Remark***: Generally, this upper bound of UR is large enough for large scale radio positioning/navigation system since the resolution of modern radio frequency synthesizer could be as low as 1Hz [11] and the corresponding UR is tens of thousands of kilometers. So the key problem left is how probable that $\{k_i\}$ being relatively prime.

*B. Probability that $\Lambda_{up}$ Is Obtained Under RSF configuration*

As revealed in lemma 1, the probability that UR under RSF set $\{f_i\}$ (i=1,…,*M*) obtains its upper bound is the same as the probability that *M* random algebraic integers $\{k_i\}$ being relatively prime. In the field of analytical number theory, a similar problem has been solved by Benkoski's theorem [12]. However, those random integers are required to be chosen from a continuous set [1,*N*]. Here, we try to tackle the difficulties arising from the arbitrary starting point $K_{min}$ and discontinuous candidate set $\mathbb{K}$ ($L \geq 1$) in RSF model.

The probability $P_M$ that integers $\{k_i\}$ are relatively prime depends not only on *M*, but also on *L* and the distribution of $\mathbb{K}$. Under the practical constraint that *L* and *M* are much smaller than *N*, we manage to specify the "growth" of $P_m$ with respect to those parameters using growth estimation technique borrowed from analytic number theory.

**Theorem 2**. The probability that *M* random algebraic integers $\{k_i\}$ out of $\mathbb{K}$ being relatively prime could be approximated as:

$$P_M \approx \frac{1}{\zeta(M)}.$$

***Proof:*** Let $p_1, p_2, \cdots$ denote distinct primes. To simplify the analysis, we assume $M \geq 3$. The following notations will be used:

$A_{p_1 \cdots p_n}$ : Number of *M*-tuple of positive integers belonging to $\mathbb{K}$ which can be divided by $\prod_{i=1}^{n} p_i$ .

$Z$ : Number of *M*-tuple of positive integers belonging to $\mathbb{K}$ which are relatively prime.

Obviously, $P_M = Z / N^M$ . Those M-tuple of integers are relatively prime if and only if there exists no prime that divides all M integers. According to the Inclusion-Exclusion principle, we have

$$Z = N^M - \sum_{p_1} A_{p_1} + \sum_{p_1 < p_2} A_{p_1 p_2} - \sum_{p_1 < p_2 < p_3} A_{p_1 p_2 p_3} + \cdots. \tag{5}$$

By using the Möbius function $\mu$, Eq. (5) could be rewritten more compactly as

$$Z = \sum_{j=1}^{\infty} \mu(j) x_j^M , \tag{6}$$

where $x_j$ denotes the number of integers in $\mathbb{K}$ which can be divided by *j* and $\mu(j)$ is a number theoretic function defined as

$$\mu(j) = \begin{cases} 1 & \text{if } j = 1 \\ 0 & \text{if } j \text{ has one or more repeated prime factors} \\ (-1)^r & \text{if } j \text{ has } r \text{ distinct primes} \end{cases}$$

It can be easily verified that $x_j^M$ is the number of *M*-tuple of positive integers in $\mathbb{K}$ that can be divided by *j*. Actually, we do not need the infinity at the upper end of summation in Eq. (6), since the terms with $j > K_{max}$ are all zeros. So we further rewrite the sum as

$$Z = \sum_{j=1}^{K_{\max}} \mu(j) x_j^M . \tag{7}$$

Let $x_j(l)$ denotes the number of integers belonging to the *l*-th segment of N which can be divided by *j*, then $x_j = \sum_{l=1}^{L} x_j(l)$. Apparently $x_j(l) = \left\lfloor \frac{N_l}{j} \right\rfloor$, so we have $\sum_{l=1}^{L}(\frac{N_l}{j} - 1) \leq x_j \leq \sum_{l=1}^{L}(\frac{N_l}{j})$, that is, $\frac{N}{j} - L \leq x_j \leq \frac{N}{j}$ since $N = \sum_{l=1}^{L} N_l$. Furthermore, *L* is assumed to be a small constant independent of *N*, we have $x_j = \boldsymbol{O}\left(\frac{N}{j}\right)$.

Since

$$x_j^M - \left(\tfrac{N}{j}\right)^M = \left(x_j - \tfrac{N}{j}\right)\left(x_j^{M-1} + x_j^{M-2}\left(\tfrac{N}{j}\right) + \cdots + \left(\tfrac{N}{j}\right)^{M-1}\right),$$

and $0 \leq \frac{N}{j} - x_j \leq L$, we obtain

$$x_j^M - \left(\tfrac{N}{j}\right)^M = O\left(\left(\frac{N}{j}\right)^{M-1}\right). \tag{8}$$

Applying this growth estimate to $Z$, we have

$$Z = \sum_{j=1}^{K_{\max}} \mu(j)\left(\tfrac{N}{j}\right)^M + O\left(\sum_{j=1}^{K_{\max}} \left(\tfrac{N}{j}\right)^{M-1}\right). \tag{9}$$

Therefore, $P_M$ could be expressed as

$$P_M = \frac{Z}{N^M} = \sum_{j=1}^{K_{\max}} \frac{\mu(j)}{j^M} + O\left( N^{-1} \sum_{j=1}^{K_{\max}} \frac{1}{j^{M-1}} \right). \tag{10}$$

As proved in [13], the summation of Dirichlet series $\mu(j)/j^M$ is $\sum_{j=1}^{\infty} \frac{\mu(j)}{j^M} = \frac{1}{\zeta(M)}$ where $\zeta(\cdot)$ is the Riemann-zeta function. So the first sum in Eq. (10) could be rewritten as:

$$\begin{aligned} \sum_{j=1}^{K_{\max}} \frac{\mu(j)}{j^M} &= \frac{1}{\zeta(M)} - \sum_{j=K_{\max}+1}^{\infty} \frac{\mu(j)}{j^M} = \frac{1}{\zeta(M)} + O\left( \int_{K_{\max}+1}^{\infty} \frac{dx}{x^M} \right) \\ &= \frac{1}{\zeta(M)} + O\left( (K_{\max}+1)^{1-M} \right), \end{aligned} \tag{11}$$

Under the condition that $N \to \infty$, $K_{max}$ also approaches to infinity. Since $M \geq 3$, we have $\mathbf{O}((K_{max}+1)^{1-M}) \to 0$.

For the second term in Eq. (10), also under the condition that $M \geq 3$ and $K_{max} \to \infty$, we have

$$\sum_{j=1}^{K_{\max}} \frac{1}{j^{M-1}} = O\left( \int_1^{K_{\max}} \frac{dx}{x^{M-1}} \right) = O(1)$$

and thus this term is on the order of $\mathbf{O}(N^{-1})$ which also approaches to 0 when $N \to \infty$.

Finally, we conclude that

$$\lim_{N\to\infty} P_M = \frac{1}{\zeta(M)} \tag{12}$$

As mentioned above, $f_0$ may be as low as $1Hz$ while the available bandwidth enabled by cognitive radios is usually up to tens of megahertz, so $N$ is a very large integer in practical scenario. Therefore, the approximation made in theorem 2 is good enough to specify the probability that those $\{k_i\}$ randomly chosen from $\mathbb{K}$ are relatively prime.

## IV. Simulation Results

In this section, we would verify the UR of MPR under RSF configuration through extensive simulations. Firstly, the probability that $\Lambda_{up}$ is obtained without phase noise is analyzed through numerical simulation. Then we show the different impact of phase noise on UR under LSF and RSF configurations and give a qualitative explanation utilizing the discrepancy function of LS range estimator.

### *A. Theoretical UR without Phase Noise*

As stated in Lemma 1, the probability that $\Lambda_{up}$ is obtained under RSF configuration is just the same as the probability $P_M$ that $M$ randomly chosen carrier frequencies being relatively prime, that is, $\kappa = 1$. Here we conduct Monte-Carlo simulations to study the relationship among $P_M$, the number of carriers $M$ and the number of segments $L$ to verify the approximation made in (13). The frequency resolution is $f_0 = 1$kHz and total bandwidth $\boldsymbol{B}$ ranges from 132MHz to 862MHz. The number of available frequencies is $N = 2^{15}$, less than 5% of total frequencies. Those frequencies follow a uniform distribution within $\boldsymbol{B}$. We examine three different scenarios where the number of spectrum segments $L$ is 1, 7, and 12 respectively. $M$ ranges from 3 to 13. The corresponding $\{k_i\}$ are chosen randomly from the available bandwidth with an M-sequence generator. For each combination of $M$ and $L$, $10^5$ Monte-Carlo simulations are conducted.

Both theoretical and simulation results about the probability $P_M$ are given in Fig. 2. As shown by the figure, simulation results highly correspond with the theoretical one in all scenarios, regardless of the number of segments $L$. Another important message delivered from the figure is that $P_M$ would approach to 1 very closely when $M > 10$, three orders less than $N$. This is very inspiring since UR could be easily extended with only a small fraction of $M$ frequencies rather than the total $N$ frequencies.

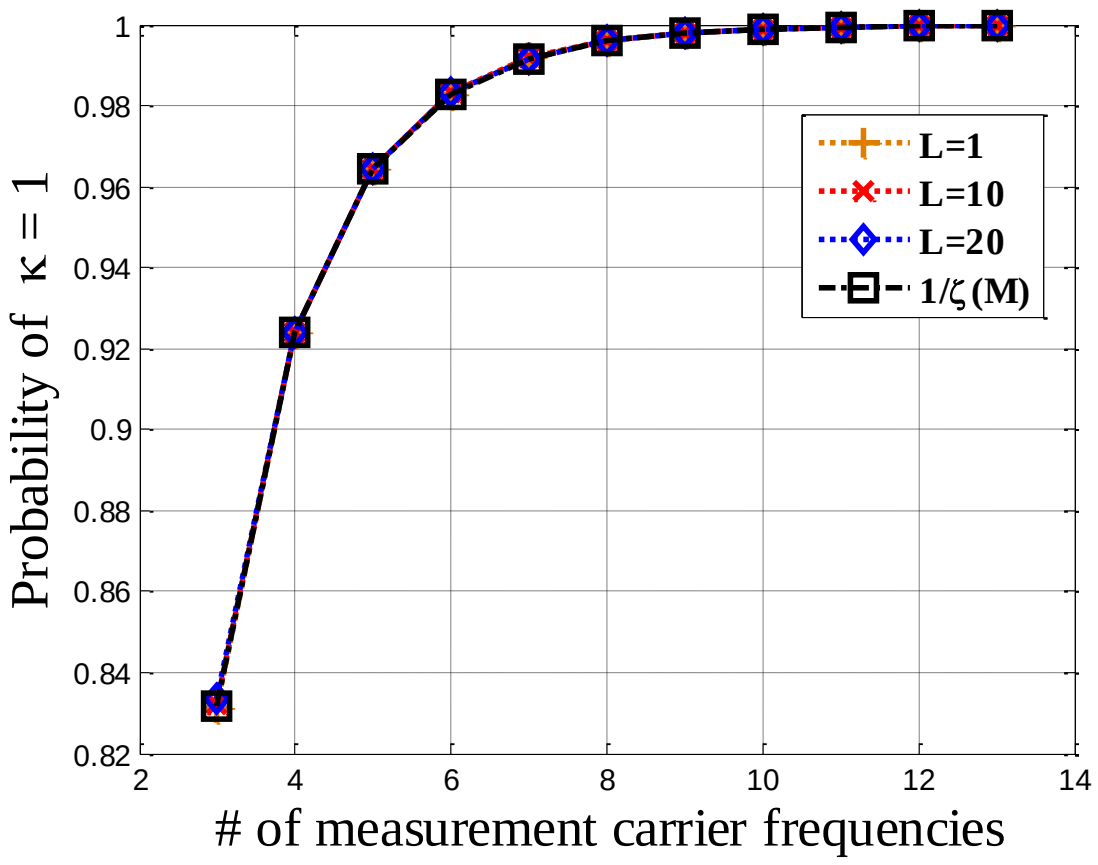

Figure 2. The probability of M positive integers $\{k_i\}$ in RSF set being relatively prime

## *B. Practical UR with Phase Noise*

Theoretically, once the condition that $\{k_i\}$ are relatively prime is met, the UR of MPR would obtain its upper bound $\Lambda_{\text{up}}$. It is noteworthy that this conclusion also holds for LSF configuration. For example, if we choose $M$ frequencies linearly from the whole available bandwidth, it is not uncommon to find that those frequencies might be relatively prime. Denote their frequency step as $\Delta f = s \cdot f_0$, then a puzzle arises: **the UR under this specific LSF configuration is either $\mathbf{C}/\boldsymbol{f_0}$ or $\mathbf{C}/\boldsymbol{\Delta f}$?**

To get some insight into this problem, we turn to the discrepancy function of LS range estimator. Here we set $f_0 = 1\text{MHz}$ and thus the corresponding upper bound of UR is $\Lambda_{\text{up}} = 300\text{m}$. Then we choose 100 and 1000 frequencies linearly with step size of 6MHz, which means the practical UR is $\Lambda = 50\text{m}$. The true range is 201 meters and the available bandwidth starts from 131.9MHz such that the chosen $\{k_i\}$ are relatively prime. The discrepancy functions under these two configurations are shown in Fig. 3(a) and 3(b) respectively.

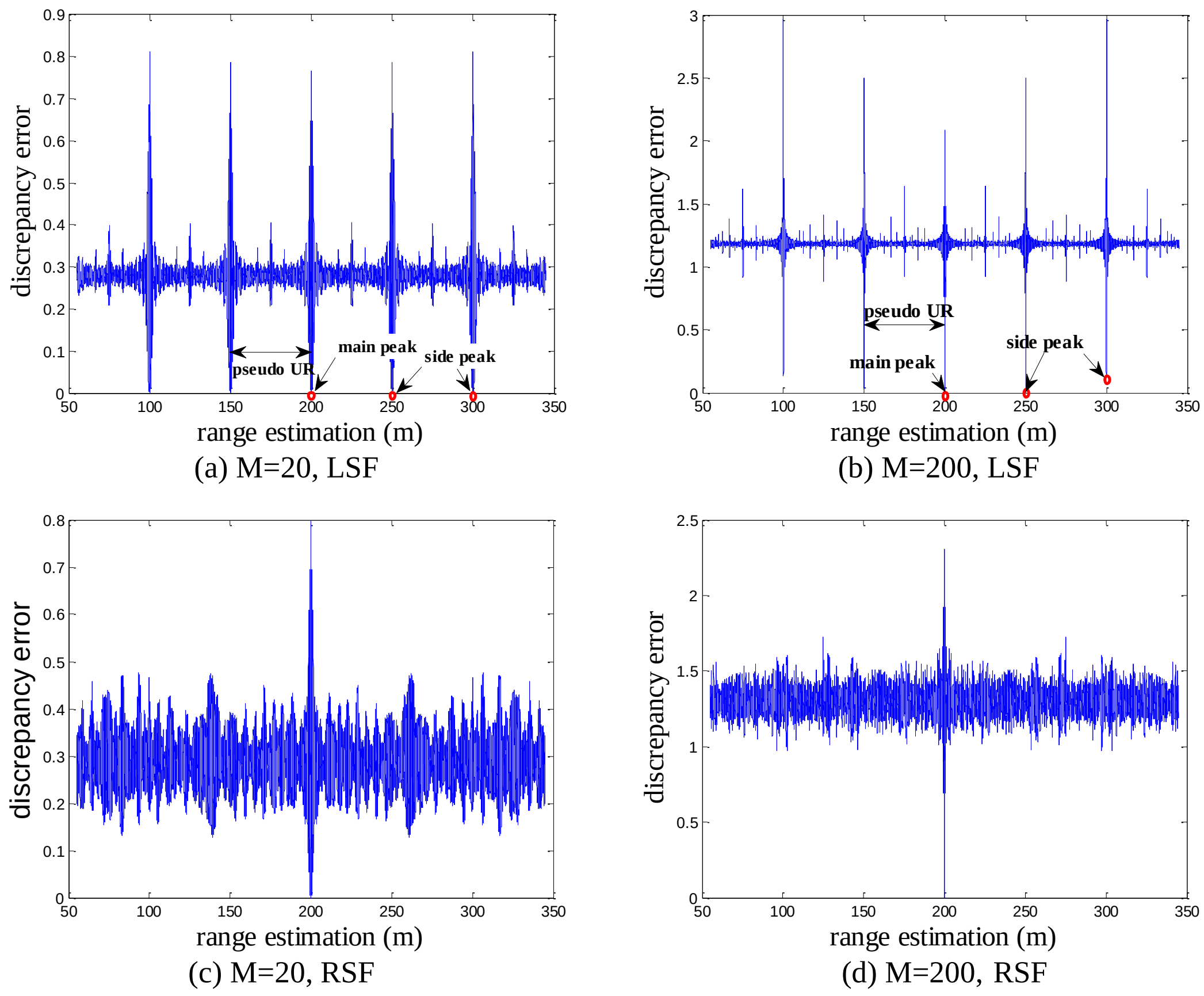

Figure 3. The discrepancy function under LSF and RSF configurations

The normal thing is that the minimal discrepancy error is located at the true range uniquely within the whole $\Lambda_{up}$, but the odd thing is there exist several side lobes around the main lobe with the interval of $\Lambda$, and their peaks are very close to that of the main lobe. Take the Fig. 3(a) for example, the difference between the main peak (d=201m) and those side peaks is far less than 0.001. After introducing random phase noise, the side peaks are no longer distinguishable from the main peak. So we define $\Lambda$ as pseudo UR which is also the practical UR adopted by engineering systems such as OMEGA and CP-GPS. The probability of pseudo UR could only be reduced by employing a sufficiently large frequency set. As shown in Fig 3(b), after M is increased from 20 to 200, the side peaks become more recognizable.

On the contrary, the discrepancy function under RSF configuration is not so sensitive to phase noise, as shown in Fig. 3(c) and 3(d). Compared with discrepancy function of LSF, there are no comparable side peaks with the main peak. Anticipatorily, the UR under RSF configuration could obtain $\Lambda_{up}$ more probable even in the presence of phase noise.

Finally, simulations are conducted to show the robustness of UR against phase noise under LSF and RSF configurations. Phase noise is modeled as i.i.d random variables with zero mean Gaussian distribution. The number of carriers $M$ ranges from 5 to 200. For each $M$, we keep the total bandwidth $\boldsymbol{B}$ of LSF and RSF configuration the same and conduct 5000 Monte-Carlo simulations to study the impact of phase noise. The results are shown in Fig. 4. Under LSF configuration, the unambiguous estimates mean those estimations located within the pseudo UR $\Lambda$. Under RSF configuration, there exists no explicit pseudo UR and a more restrict criterion is used: $\mathrm{C}/(\Delta f)_{max}$ where $(\Delta f)_{max}$ is the maximum step size. Monte-Carlo simulations are conducted on 1000 realizations of RSF sequence and the result within 95% confidential interval is adopted.

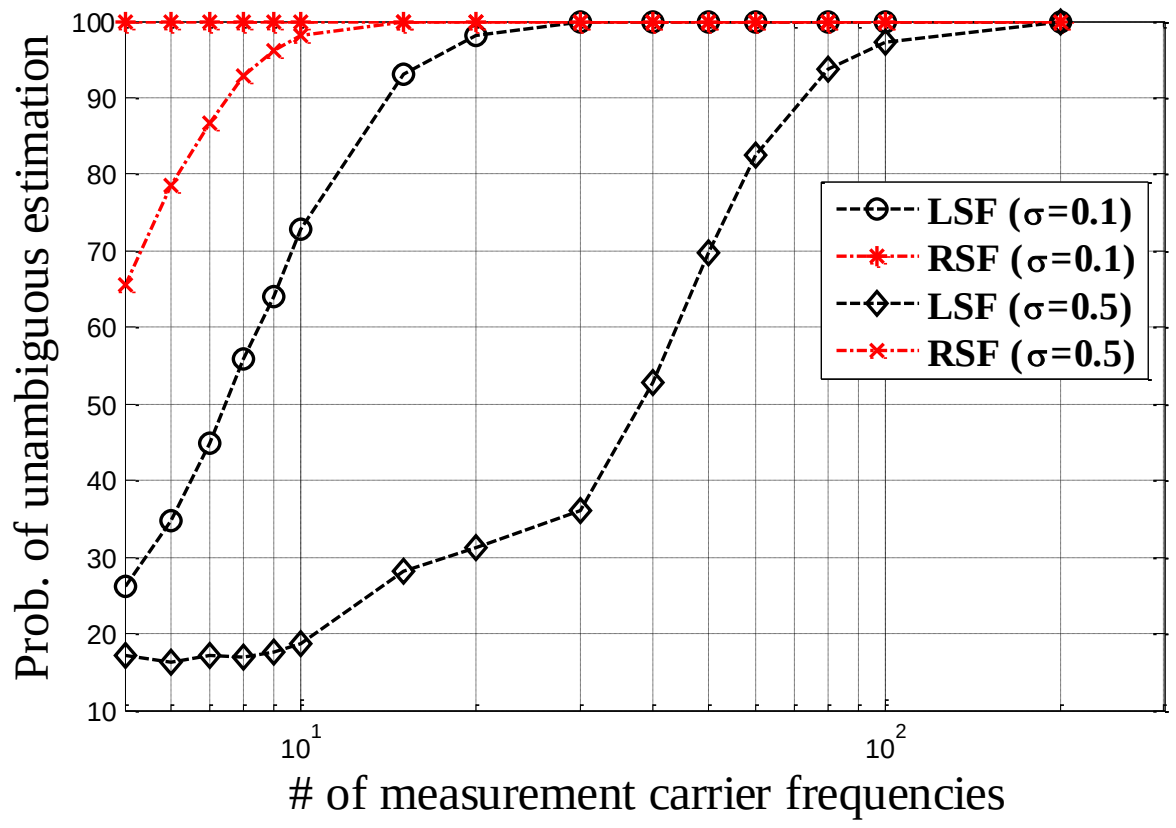

Figure 4. The probability of unambiguous estimation of LSF and RSF with phase noise

As shown in Fig. 4, under RSF configuration, the probability of unambiguous estimation within the $\Lambda_{up}$ in the case of small phase error ($\sigma = 0.1$) still follows the theoretical result in Fig. 2. When phase error is increased ($\sigma = 0.5$), this probability drops when $M$ is small but increases to 1 rapidly after $M > 20$. On the contrast, the UR under LSF configuration could only obtain its $\Lambda_{up}$ with high probability by employing tens of or even hundreds of carriers. Fig. 5 gives a more thorough explanation of this phenomenon. In the presence of phase noise, the probability that range estimation under LSF configuration locates within each pseudo UR $\Lambda$ is almost the same until $M$ is sufficiently large.

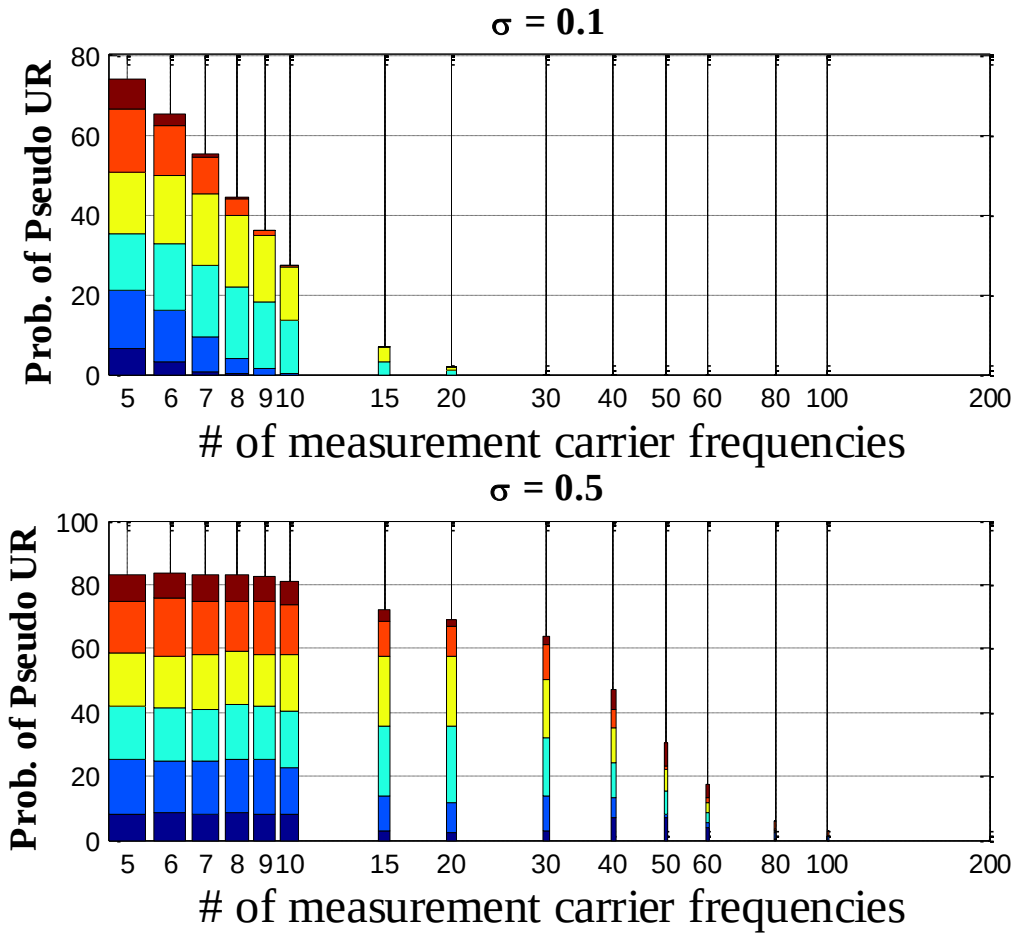

Figure 5. The probability of pseudo UR in LSF after phase noise is introduced

## V. CONCLUSIONS

The unambiguous range of MPR under random carrier sequence has not been explored. We make a first attempt to describe its two important stochastic properties: (i) the upper bound of UR under RSF configuration $\Lambda_{\mathrm{up}}$; and (ii) the probability that $\Lambda_{\mathrm{up}}$ could be obtained $P_M$. It is inspiring to find that $\Lambda_{\mathrm{up}}$ under RSF configuration may be readily extended to that under LSF configuration. Furthermore, this upper bound could be obtained with very high probability when only a dozen of carriers are employed, as long as the number of candidate frequencies $\boldsymbol{N}$ is large enough. This conclusion holds no matter how many segments the bandwidth are separated into or how the segments are distributed.

We further study the robustness of UR against phase noise and uncover the existence of multiple pseudo UR within $\Lambda_{\mathrm{up}}$ under LSF configuration. On the contrary, the UR under RSF configuration is more resilient against phase noise, since the randomness of frequencies disrupts the pace of pseudo UR. Therefore, the theoretical UR $\Lambda_{\mathrm{up}}$ is also obtainable in practical scenario. Our work eliminates the fundamental tradeoff between the UR and real time performance of MPR and facilitates the application of MPR in large scale radio positioning/navigation system.

Some issues are still left for future research. Firstly, although a small RSF set is enough to scale UR, it may sacrifice the ranging accuracy. So both the UR and accuracy should be taken into consideration when determining the number of measurement frequencies. Secondly, the UR is also dependent on range estimator. We focus on UR with LS estimator in this paper while its property with other estimators should be further explored.

## ACKNOWLEDGEMENT

This work was supported by the National Natural Science Foundation of China (61273047, 61402520, 61573376 and 61071115). Wangdong Qi is the corresponding author.

We would like to thank the anonymous reviewers for their constructive comments and Prof. Xinbing Wang, Dr. Luoyi Fu for their valuable suggestions.